\documentclass[pra,superscriptaddress]{revtex4}% Physical Review B
\usepackage{graphicx}% Include figure files
\usepackage{dcolumn}% Align table columns on decimal point
\usepackage{bm}% bold math
\usepackage{amsmath}
\usepackage{amssymb,amsbsy}
\usepackage{color,ulem}% 
\newcommand{\beq}{\begin{equation}}
\newcommand{\eeq}{\end{equation}}
\newcommand{\bea}{\begin{eqnarray}}
\newcommand{\eea}{\end{eqnarray}}
\newcommand{\bec}{\begin{center}}
\newcommand{\enc}{\end{center}}
\newcommand{\bfr}{\begin{flushright}}
\newcommand{\efr}{\end{flushright}}
\newcommand{\la}{\langle}
\newcommand{\ra}{\rangle}

\newcommand{\alp}{\alpha}

\newcommand{\om}{\omega}
\newcommand{\kap}{\kappa}

\newcommand{\sig}{\sigma}

\newcommand{\Om}{\Omega}

\newcommand{\tom}{\widetilde{\omega}}

\newcommand{\tkap}{\widetilde{\kappa}}
\newcommand{\tone}{\widetilde{1}}
\newcommand{\ttwo}{\widetilde{2}}
\newcommand{\tthree}{\widetilde{3}}
\newcommand{\tfour}{\widetilde{4}}

\newcommand{\cF}{{\cal F}}
\newcommand{\cH}{{\cal H}}

\begin{document}
%%%%%%%%%%%%%%%%%%%%%%%%%%%%%%%%%%%%%%%%%%%%%%%%%%%%%%%%%%%%%%%%%%%%%%%%%%%%%%%
%%%%%%%%%%%%%%%%%%%%%%%%%%%%%%%%%%%%%%%%%%%%%%%%%%%%%%%%%%%%%%%%%%%%%%%%%%%%%%%
\title{Tunable quantum gate between a superconducting atom \\ 
and a propagating microwave photon
}
%%%%%%%%%%%%%%%%%%%%%%%%%%%%%%%%%%%%%%%%%%%%%%%%%%%%%%%%%%%%%%%%%%%%%%%%%%%%%%%
%%%%%%%%%%%%%%%%%%%%%%%%%%%%%%%%%%%%%%%%%%%%%%%%%%%%%%%%%%%%%%%%%%%%%%%%%%%%%%%
\author{K. Koshino}
\affiliation{College of Liberal Arts and Sciences, Tokyo Medical and Dental
University, Ichikawa, Chiba 272-0827, Japan}
\author{K. Inomata}
\affiliation{RIKEN Center for Emergent Matter Science (CEMS), 2-1 Hirosawa, Wako, 
Saitama 351-0198, Japan}
\author{Z. R. Lin}
\affiliation{RIKEN Center for Emergent Matter Science (CEMS), 2-1 Hirosawa, Wako, 
Saitama 351-0198, Japan}
\author{Y. Tokunaga}
\affiliation{NTT Secure Platform Laboratories, NTT Corporation, Musashino 180-8585, Japan}
\author{T. Yamamoto}
\affiliation{IoT Device Research Laboratories, NEC Corporation, Tsukuba, Ibaraki 305-8501, Japan}
\author{Y. Nakamura}
\affiliation{RIKEN Center for Emergent Matter Science (CEMS), 2-1 Hirosawa, Wako, 
Saitama 351-0198, Japan}
\affiliation{Research Center for Advanced Science and Technology (RCAST), 
The University of Tokyo, Meguro-ku, Tokyo 153-8904, Japan}

%%%%%%%%%%%%%%%%%%%%%%%%%%%%%%%%%%%%%%%%%%%%%%%%%%%%%%%%%%%%%%%%%%%%%%%%%%%%%%%
%%%%%%%%%%%%%%%%%%%%%%%%%%%%%%%%%%%%%%%%%%%%%%%%%%%%%%%%%%%%%%%%%%%%%%%%%%%%%%%
\date{\today}
%%%%%%%%%%%%%%%%%%%%%%%%%%%%%%%%%%%%%%%%%%%%%%%%%%%%%%%%%%%%%%%%%%%%%%%%%%%%%%%
%%%%%%%%%%%%%%%%%%%%%%%%%%%%%%%%%%%%%%%%%%%%%%%%%%%%%%%%%%%%%%%%%%%%%%%%%%%%%%%
\begin{abstract}
We propose a two-qubit quantum logic gate 
between a superconducting atom and a propagating microwave photon. 
The atomic qubit is encoded on its lowest two levels
and the photonic qubit is encoded on its carrier frequencies. 
The gate operation completes deterministically upon reflection of a photon, 
and various two-qubit gates 
(SWAP, $\sqrt{\rm SWAP}$, and Identity) are realized
through {\it in situ} control of the drive field. 
The proposed gate is applicable to construction of 
a network of superconducting atoms, 
which enables gate operations between non-neighboring atoms. 
% are connected quantum-mechanically by single microwave photons.  
\end{abstract}
%%%%%%%%%%%%%%%%%%%%%%%%%%%%%%%%%%%%%%%%%%%%%%%%%%%%%%%%%%%%%%%%%%%%%%%%%%%%%%%
% \pacs{
% 42.50.Ar, % Photon statistics and coherence theory 
% 42.50.Pq, % Cavity quantum electrodynamics; micromasers 
% 42.65.Sf % optical spatio-temporal dynamics 
% }
\maketitle

Physical implementation of a scalable quantum system 
that enables quantum computation 
is one of the main objectives in modern quantum technology. 
There are two approaches for achieving this goal. 
In the first approach, we construct a quantum circuit 
which is composed of qubits of the same kind:
the one-qubit gates are realized 
by local operations on a single qubit, 
and the two-qubit gates are realized 
by mutual interaction between a pair of qubits. 
For example, high-fidelity gate operations reaching 
the fault tolerance threshold for surface code error correction~\cite{surface} 
have been achieved in an array of superconducting qubits~\cite{Xmon}. 
Recently, a scalable Shor's algorithm~\cite{Kita} has been demonstrated
using a trapped ion quantum computer~\cite{blatt}.

In the second approach, 
which is known as the distributed or modular architecture, 
we use a hybrid quantum network composed of flying and stationary 
qubits~\cite{dis1,dis2,Kim,Ben,mod1,mod2}. 
Flying qubits, which are typically implemented by photons, 
transfer quantum information among the stationary nodes. 
The stationary qubits, which are implemented by real or artificial atoms,
are used to register and process quantum information. 
Construction of such hybrid quantum networks has been developed 
actively in cavity quantum electrodynamics (QED) 
using real atoms and optical photons. 
For example, a deterministic quantum gate between 
a propagating photon and an atom has been demonstrated,
which has been further extended to a photon-photon gate~\cite{duan,Rem2,Rem4}. 
The observation of single-photon Raman interaction~\cite{sprint1,sprint2}
would be a crucial step towards achieving 
the swap-based photon-photon gates~\cite{rswap}.
Similarly, in the microwave quantum-optics setups 
based on circuit QED~\cite{cQED1,cQED2}, 
we can connect superconducting atoms by microwave photons propagating in waveguides.
Recently, entanglement generation between two remote superconducting atoms 
has been achieved by interfering the two microwave photons emitted by the atoms~\cite{Yale}.

\begin{figure}[t]
\begin{center}
\includegraphics[width=8cm]{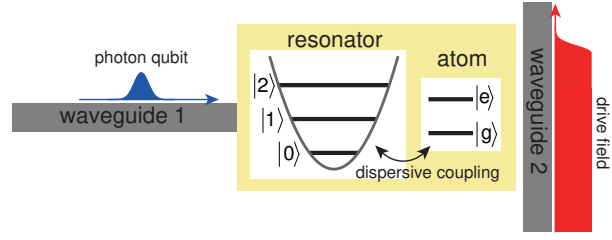}
\caption{
Schematic of the tunable atom-photon quantum gate.
We input a photon qubit through waveguide~1
and drive the superconducting atom through waveguide~2. 
The quantum-gate operation completes upon reflection of the photon. 
We can realize various types of quantum gate by changing the drive condition. 
}
\label{fig:sch}
\end{center}
\end{figure}

In this study, we propose a new scheme for 
implementing deterministic two-qubit gates 
between a superconducting atom and a propagating microwave photon. 
In the proposed device, a driven superconducting atom 
is coupled to a waveguide photon via a resonator (Fig.~\ref{fig:sch}). 
The atomic qubit is encoded on its two lowest levels ($|g\ra$ and $|e\ra$), and 
the photon qubit is encoded on its carrier frequencies~\cite{freq_qubit}. 
The gate operation completes deterministically upon reflection of a photon. 
A remarkable feature of the proposed gate is its tunability: 
through {\it in situ} control of the drive field to the atom, 
we can continuously change the gate type,
including SWAP, $\sqrt{\rm SWAP}$, and Identity gates
which are of practical importance. 
Furthermore, by cascading the proposed devices, 
we can execute an entangling % a $\sqrt{\rm SWAP}$ 
gate between two remote superconducting atoms. 
This implies the realization of a universal gate set, 
since one-qubit gate operations are easy in superconducting atoms. 

The schematic of the considered device is shown in Fig.~\ref{fig:sch}.
A superconducting artificial atom, 
which can be regarded as a two-level system,
is dispersively coupled to a resonator. 
The resonator and the atom are respectively coupled to waveguides~1 and 2.
Through waveguide~1, we input a microwave-photon qubit,
whose quantum information is encoded on its carrier frequencies.
Through waveguide~2, we apply a drive field to the atom 
in order to engineer the dressed states of the atom-resonator system~\cite{dse}.
Assuming a static drive field of amplitude $\Om_d$ and frequency $\om_d$, 
the Hamiltonian of the atom-resonator system is given, 
in the rotating frame, by
\bea
\cH_{ar} &=& 
\om_r a^{\dag}a\sig\sig^{\dag} + 
[(\om_a-\om_d)+(\om_r-2\chi)a^{\dag}a]\sig^{\dag}\sig + \Om_d(\sig^{\dag}+\sig),
\label{eq:Har}
\eea
where $\sig$ ($a$) is the annihilation operator for the atom (resonator),
$\om_a$ ($\om_r$) is the resonance frequency of the atom (resonator),
and $\chi$ is the dispersive shift.
For concreteness, we assume the following parameter values:
$\om_a/2\pi=5$~GHz, $\om_r/2\pi=10$~GHz, and $\chi/2\pi=75$~MHz.

\begin{figure}[t]
\begin{center}
\includegraphics[width=8cm]{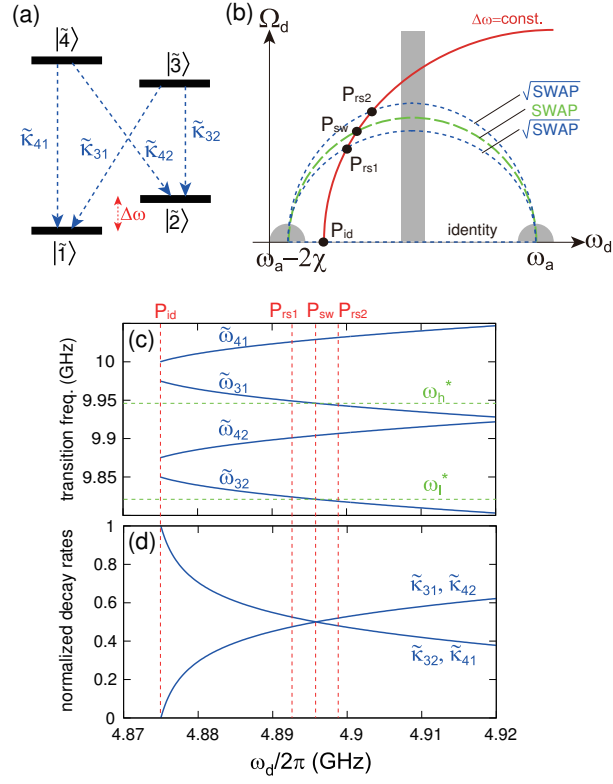}
\caption{
Dressed-state engineering.
(a)~Level structure of the dressed states in the rotating frame.
(b)~Drive conditions to achieve various quantum gates. 
We change the drive condition along the red solid line, 
where $\Delta\om=\tom_{21}$ is kept constant. 
SWAP gate is realized at $\mathrm{P}_\mathrm{sw}$, 
$\sqrt{\mathrm{SWAP}}$ is realized at $\mathrm{P}_\mathrm{rs1/2}$, and 
Identity gate is realized at $\mathrm{P}_\mathrm{id}$.
In the shadowed areas, the gate fidelities are degraded due to the parasitic excitations. 
(c)~Transition frequencies $\tom_{ij}$ and 
(d)~normalized decay rates $\tkap_{ij}/\kap$ as functions of $\om_d$. 
$\Om_d$ is adjusted to satisfy $\Delta\om/2\pi=125$~MHz. 
}
\label{fig:dse}
\end{center}
\end{figure}

Throughout this study, 
we use the lowest four levels of the atom-resonator system,
$|g,0\ra$, $|e,0\ra$, $|g,1\ra$, and $|e,1\ra$. 
These {\it bare} states are the eigenstates of $\cH_{ar}$
when the drive field is off ($\Om_d=0$). 
We set the drive frequency $\om_d$ within the range of $\om_a-2\chi<\om_d<\om_a$.
Then, in the frame rotating at $\om_d$, 
we obtain a nested energy diagram of the bare states, where
$\om_{|g,0\ra}<\om_{|e,0\ra}<\om_{|e,1\ra}<\om_{|g,1\ra}$. 
When the drive field is on,
the bare states are hybridized to form the dressed states. 
We label them from the lowest in energy and 
denote them by $|\tone\ra$, $|\ttwo\ra$, $|\tthree\ra$, and $|\tfour\ra$
[Fig.~\ref{fig:dse}(a)].
Diagonalizing $\cH_{ar}$, they are given by
\bea
|\tone\ra &=& \cos\theta_{l}|g,0\ra - \sin\theta_{l}|e,0\ra,\label{eq:tone}
\\
|\ttwo\ra &=& \sin\theta_{l}|g,0\ra + \cos\theta_{l}|e,0\ra,\label{eq:ttwo}
\\
|\tthree\ra &=& \cos\theta_{h}|e,1\ra - \sin\theta_{h}|g,1\ra,
\\
|\tfour\ra &=& \sin\theta_{h}|e,1\ra + \cos\theta_{h}|g,1\ra,\label{eq:tfour}
\eea
where 
$\theta_{l} = \frac{1}{2}\mathrm{arg}(\frac{\om_a-\om_d}{2}+i\Om_d)$ and 
$\theta_{h} = \frac{1}{2}\mathrm{arg}(\frac{\om_d-\om_a+2\chi}{2}+i\Om_d)$. 
% $\theta_{l} = \frac{1}{2}\tan^{-1}(\frac{2\Om_d}{\om_a-\om_d})$ and 
% $\theta_{h} = \frac{1}{2}\tan^{-1}(\frac{2\Om_d}{\om_d-\om_a+2\chi})$. 
Their eigenenergies are given by
\bea
\tom_{1,2} &=& 
\textstyle{
\frac{\om_a-\om_d}{2} \pm \sqrt{\left(\frac{\om_a-\om_d}{2}\right)^2+\Om_d^2},
}
\label{eq:tom12}
\\
\tom_{3,4} &=& 
\textstyle{
\om_r-\frac{\om_d-\om_a+2\chi}{2} \pm \sqrt{\left(\frac{\om_d-\om_a+2\chi}{2}\right)^2+\Om_d^2},
}
\eea
where the plus (minus) sign is taken for $\tom_2$ and $\tom_4$ ($\tom_1$ and $\tom_3$). 
In this four level system, 
$|\tthree\ra$ and $|\tfour\ra$ decay to $|\tone\ra$ and $|\ttwo\ra$ 
emitting a photon into waveguide~1.
Denoting the radiative decay rate of resonator by $\kappa$, 
the decay rates between the dressed states are given by
\bea
\tkap_{32}=\tkap_{41}&=&\kap\cos^2\theta_t,
\label{eq:1om}
\\
\tkap_{31}=\tkap_{42}&=&\kap\sin^2\theta_t,
\label{eq:2om}
\eea
where $\theta_t=\theta_l+\theta_h$.

We discuss the response of this four-level system
to a single microwave photon input through waveguide~1. 
For simplicity, we assume 
that the input photon is monochromatic with frequency $\om$. 
Furthermore, we assume that both $|\tone\ra$ and $|\ttwo\ra$ are stable
and the four-level system is in their superposition initially. 
Due to the oblique decay paths ($\tkap_{31}$ and $\tkap_{42}$), 
the input photon may induce the Raman transition upon reflection. 
The state vector of the overall system, 
consisting of a propagating photon and the dressed states, evolves as
\bea
|\tone,\om\ra & \to & \xi_{11}(\om)|\tone,\om\ra + \xi_{12}(\om)|\ttwo,\om-\Delta\om\ra,
\\
|\ttwo,\om\ra & \to & \xi_{21}(\om)|\tone,\om+\Delta\om\ra + \xi_{22}(\om)|\ttwo,\om\ra,
\eea
where $\Delta\om=\tom_{21}=\tom_2-\tom_1$.
The coefficients $\xi_{ij}$ are given by (see Appendix~\ref{sec:A})
\bea
\xi_{11}(\om) &=& 
1-\frac{\kap\sin^2 \theta_t}{\frac{\kap}{2}-i(\om-\tom_{31})}
 -\frac{\kap\cos^2 \theta_t}{\frac{\kap}{2}-i(\om-\tom_{41})},
\label{eq:xi11}
\\
\xi_{12}(\om) &=& 
 \frac{\kap\sin\theta_t\cos\theta_t}{\frac{\kap}{2}-i(\om-\tom_{31})}
-\frac{\kap\sin\theta_t\cos\theta_t}{\frac{\kap}{2}-i(\om-\tom_{41})}, 
\label{eq:xi12}
\\
\xi_{21}(\om) &=& 
 \frac{\kap\sin\theta_t\cos\theta_t}{\frac{\kap}{2}-i(\om-\tom_{32})}
-\frac{\kap\sin\theta_t\cos\theta_t}{\frac{\kap}{2}-i(\om-\tom_{42})}, 
\label{eq:xi21}
\\
\xi_{22}(\om) &=& 
1-\frac{\kap\cos^2 \theta_t }{\frac{\kap}{2}-i(\om-\tom_{32})}
 -\frac{\kap\sin^2 \theta_t }{\frac{\kap}{2}-i(\om-\tom_{42})}.
\label{eq:xi22}
\eea
We can confirm the probability conservation, 
$|\xi_{11}|^2+|\xi_{12}|^2=|\xi_{21}|^2+|\xi_{22}|^2=1$.

In the proposed atom-photon gate,  
we use $|\tone\ra$ and $|\ttwo\ra$ as the logical basis for the material node. 
Note that these states are roughly the atomic ground and excited states
($|\tone\ra\approx|g,0\ra$ and $|\ttwo\ra\approx|e,0\ra$) 
under our choice of the drive condition.
For the photonic qubit, 
we encode quantum information on its career frequency: 
the basis states are $|\om_l\ra$ and $|\om_h\ra$, 
where $(\om_l, \om_h)=(\tom_{32}, \tom_{31})$ or $(\tom_{42}, \tom_{41})$.
For concreteness, we focus on the former case
and use $|\tone\ra$, $|\ttwo\ra$ and $|\tthree\ra$ as a $\Lambda$ system hereafter.
The case of an ^^ ^^ impedance-matched'' $\Lambda$ system,
where $\theta_t=\pi/4$ and therefore $\tkap_{31}=\tkap_{32}$, 
is of particular importance. 
If $\om_l(=\tom_{32})$ is detuned sufficiently 
from the non-target transitions ($\tom_{31}$, $\tom_{41}$, and $\tom_{42}$), 
we immediately observe in Eqs.~(\ref{eq:xi11})--(\ref{eq:xi22}) that
$\xi_{11}(\om_l)=\xi_{21}(\om_l)=1$ and $\xi_{12}(\om_l)=\xi_{22}(\om_l)=0$,
which implies that 
$|\tone,\om_l\ra \to |\tone,\om_l\ra$ and $|\ttwo,\om_l\ra \to |\tone,\om_h\ra$. 
Similarly, $|\tone,\om_h\ra \to |\ttwo,\om_l\ra$ and $|\ttwo,\om_h\ra \to |\ttwo,\om_h\ra$. 
These four time evolutions are summarized as
\bea
(\alp_1|\tone\ra + \alp_2|\ttwo\ra)\otimes(\beta_1|\om_l\ra + \beta_2|\om_h\ra)
\to
% \nonumber \\
(\beta_1|\tone\ra + \beta_2|\ttwo\ra)\otimes(\alp_1|\om_l\ra + \alp_2|\om_h\ra).
\eea
where $\alp_1$, $\alp_2$, $\beta_1$ and $\beta_2$ are arbitrary coefficients.
Namely, SWAP gate is achieved between the photon and atom qubits. 
Note that the deterministic Raman transition, 
$|\tone,\om_h\ra \to |\ttwo,\om_l\ra$, 
has been demonstrated recently as the deterministic down-conversion 
and is applied for detection of single microwave photons~\cite{ino1,ino2}.

The frequency $\om_d$ and the amplitude $\Om_d$ of the qubit drive are chosen as follows: 
(i)~In order to constitute an impedance-matched $\Lambda$ system
($\theta_t=\pi/4$), $\om_d$ and $\Om_d$ should satisfy
\beq
4\Om_d^2 = (\om_a-\om_d)(\om_d-\om_a+2\chi).
\label{eq:cond2}
\eeq
This is represented as an ellipse on the $(\om_d, \Om_d)$ plane 
[green dashed line in Fig.~\ref{fig:dse}(b)].
(ii)~$\om_l(=\tom_{32})$ and $\om_h(=\tom_{31})$ should be 
detuned sufficiently from the non-target transitions. 
This requires that $(\om_d, \Om_d)\neq(\om_a-2\chi, 0)$, $(\om_d, \Om_d)\neq(\om_a, 0)$,
and $\om_d\neq\om_a-\chi$ [shadowed areas in Fig.~\ref{fig:dse}(b)]. 
(iii)~The frequency difference between the two basis states, 
$\Delta\om=\om_h-\om_l$ is given, from Eq.~(\ref{eq:tom12}), by
\beq
\Delta\om = \sqrt{(\om_a-\om_d)^2+4\Om_d^2}.
\label{eq:cond1}
\eeq
The condition that $\Delta\om=\mathrm{constant}$ 
is also represented as an ellipse on the $(\om_d, \Om_d)$ plane
[red solid line in Fig.~\ref{fig:dse}(b)].
Practically, a large $\Delta\om$ is advantageous, 
since we can suppress the effects of finite qubit lifetime
by using a short photon pulse. 
Hereafter we set $\Delta\om/2\pi=125$~MHz. 
From Eqs.~(\ref{eq:cond2}) and (\ref{eq:cond1}), 
the drive condition to achieve a SWAP gate is determined as
\bea
\om_{d}^\mathrm{sw} &=&
\textstyle{
\om_a-\frac{(\Delta\om)^2}{2\chi},
}
\\
\Om_{d}^\mathrm{sw} &=&
\textstyle{
\frac{\Delta\om}{4\chi}\sqrt{4\chi^2-(\Delta\om)^2},
}
\eea
which amount to $\om_d^\mathrm{sw}/2\pi=4.896$~GHz
and $\Om_d^\mathrm{sw}/2\pi=34.55$~MHz, respectively
[P$_\mathrm{sw}$ in Fig.~\ref{fig:dse}(b)]. 
With this qubit drive, 
the carrier frequencies of the photon qubit are determined as
\bea
\om^*_l &=& 
\textstyle{
\om_r-\chi-\sqrt{\chi^2-(\frac{\Delta\om}{2})^2}-\frac{\Delta\om}{2},
}
\\
\om^*_h &=& 
\textstyle{
\om_r-\chi-\sqrt{\chi^2-(\frac{\Delta\om}{2})^2}+\frac{\Delta\om}{2},
}
\eea
which amounts to $\om^*_l/2\pi=9.821$~GHz and 
$\om^*_h/2\pi=9.946$~GHz, respectively [Fig.~\ref{fig:dse}(c)].

A merit of the present scheme is that
the transition frequencies and the decay rates between the dressed states
are controllable through the drive field.
In particular, we can vary the drive condition 
conserving the frequency difference $\Delta\om$ between $|\tone\ra$ and $|\ttwo\ra$
[solid line in Fig.~\ref{fig:dse}(b)]. 
By changing the drive condition smoothly with a transit time of the order of 10~ns, 
we can suppress the non-adiabatic transition between $|\tone\ra$ and $|\ttwo\ra$.
This implies that various atom-photon gates can be realized 
without changing the logical basis. 
For example, when the qubit drive is off [P$_\mathrm{id}$ of Fig.~\ref{fig:dse}(b)], 
$\Om_d=0$ and therefore $\theta_t=0$. 
Then we realize an Identity gate, 
where the atom and photon qubits remain unchanged upon reflection. 
Furthermore, under different drive conditions 
[P$_\mathrm{rs1}$ and P$_\mathrm{rs2}$ of Fig.~\ref{fig:dse}(b)],
we realize a $\sqrt{\mathrm{SWAP}}$ gate, 
which generates maximal entanglement between the atom and photon qubits~\cite{sqrtswap}.
The basis states evolve as 
$|\tone,\om^*_l\ra \to |\tone,\om^*_l\ra$,
$|\tone,\om^*_h\ra \to \frac{1\mp i}{2}|\tone,\om^*_h\ra+\frac{1\pm i}{2}|\ttwo,\om^*_l\ra$,  
$|\ttwo,\om^*_l\ra \to \frac{1\pm i}{2}|\tone,\om^*_h\ra+\frac{1\mp i}{2}|\ttwo,\om^*_l\ra$, and 
$|\ttwo,\om^*_h\ra \to |\ttwo,\om^*_h\ra$,
where the upper (lower) signs should be taken at P$_\mathrm{rs1}$ (P$_\mathrm{rs2}$).

\begin{figure}[t] 
\begin{center}
\includegraphics[width=80mm]{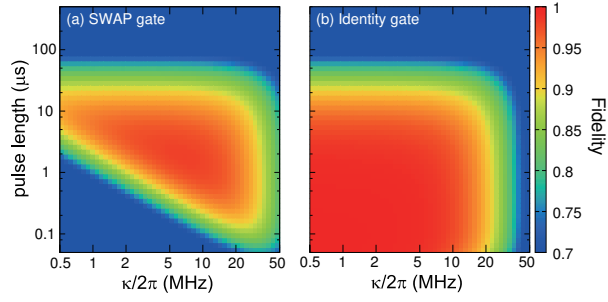}
\caption{
Average gate fidelities for (a)~SWAP and (b)~Identity gates
as functions of the linewidth $\kap$ of the resonator 
and the pulse length $l$. 
}
\label{fig:fid}
\end{center}
\end{figure}

In the above discussions, the lifetime $T_1$ of the superconducting atom 
and the length $l$ of the photon pulse are assumed to be infinite. 
Here, taking account of their finiteness, we evaluate the gate fidelity quantitatively. 
We assume a long-lived superconducting atom with $T_1=80$~$\mu$s 
and with negligible pure dephasing, and 
employ a trigonometric pulse profile for the photon qubit, as given by
\beq
f_{\om, l}(t) = % \cN \times 2^{-(2t/l)^2}\exp(-i\om t), 
\begin{cases}
\sqrt{2/l}\cos(\pi t/l)\exp(-i\om t) & (|t|<l/2)
\\
0 & ({\rm otherwise}),
\end{cases}
\label{eq:foml}
\eeq
where $\om=\om^*_l$ or $\om^*_h$. 
Note that a pulse-shaped single photon 
is available in the microwave domain~\cite{ETH}.
% In order to avoid the overlap between $|\om^*_l\ra$ and $|\om^*_h\ra$ in the frequency space, 
% the pulse length should be longer than $l \gtrsim$100~ns,
% by which $|\la\om^*_l|\om^*_h\ra|^2 \lesssim 10^{-4}$. 
For $\Delta\om/2\pi=125$~MHz, 
by choosing the pulse length $l \gtrsim 50$~ns, 
the overlap between $|\om_h\ra$ and $|\om_l\ra$ in the frequency space 
becomes negligible ($|\la\om^*_l|\om^*_h\ra| \lesssim 10^{-3}$). 
Setting the initial moment at $t=-l/2$,
we evaluate an averaged gate fidelity 
after photon reflection at $t=l/2$. 
In Fig.~\ref{fig:fid}, 
the average gate fidelities of SWAP and Identity gates are plotted 
as functions of $\kappa$ and $l$ (see Appendix~\ref{sec:B}). 
The conditions for high-fidelity SWAP gate are 
(i)~the gate time $l$ is much shorter than the lifetime $T_1$ of the atom, 
(ii)~the delay of the photon pulse due to absorption and reemission ($\sim\kap^{-1}$)
is much smaller than the pulse length $l$,
and (iii)~levels $|\tthree\ra$ and $|\tfour\ra$ are well resolved in frequency,
which requires $\kap \ll \tom_{43} \simeq 2\pi \times 70$~MHz [see Fig.~\ref{fig:dse}(c)]. 
On the other hand, 
the conditions for high-fidelity Identity gate are (i) and 
(iv)~the carrier frequencies $\om^*_l$ and $\om^*_h$ are detuned 
sufficiently from $\tom_{32}$ and $\tom_{41}$, 
which requires $\kap \ll |\tom_{32}-\om^*_l| \simeq 2\pi \times 30$~MHz [see Fig.~\ref{fig:dse}(c)]. 
By setting $\kap/2\pi=5.236$~MHz and $l=1.738~\mu$s, 
the gate fidelities reach $\cF_\mathrm{id}=0.986$,
$\cF_\mathrm{sw}=0.980$, $\cF_\mathrm{rs1}=0.986$, and $\cF_\mathrm{rs2}=0.986$.
These fidelities are sufficient for 
the communication channel in the distributed architecture~\cite{Ben}.
We can further improve the gate fidelities 
by enhancing the lifetime $T_1$ of the atom and the dispersive shift $\chi$. 
%---------------------------------------------------
%---------------------------------------------------

%---------------------------------------------------
\begin{figure}[t] 
\begin{center}
\includegraphics[width=75mm]{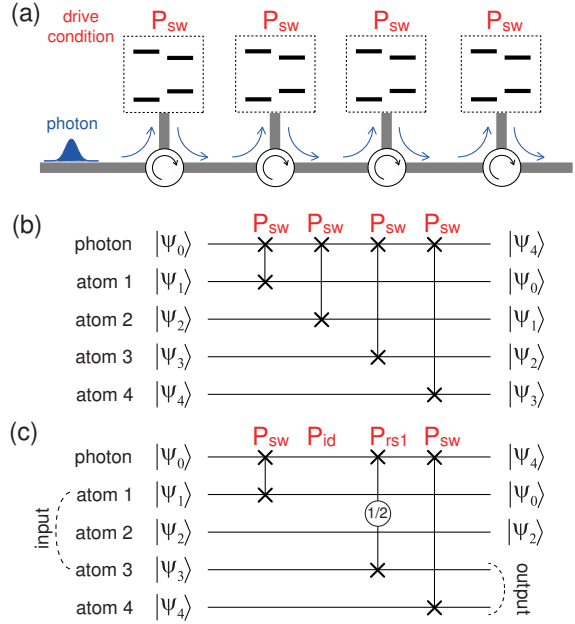}
\caption{
One-dimensional quantum circuit. 
(a)~Schematic of the circuit. 
Superconducting atoms are cascaded by circulators. 
The drive conditions of the atoms are individually controllable. 
(b)~Circuit diagram of the quantum domino. 
(c)~Circuit diagram of the atom-atom $\sqrt{\rm SWAP}$ gate. 
The input qubits are atoms~1 and 3, and the output qubits are atoms~3 and 4.
}
\label{fig:1D}
\end{center}
\end{figure}
%---------------------------------------------------

By cascading such atom-resonator systems 
using circulators [Fig.~\ref{fig:1D}(a)], 
we build up a one-dimensional network of atomic qubits
which are connected quantum-mechanically by propagating photons. 
For example, we present the circuit diagram of 
a ^^ ^^ quantum domino'' in Fig.~\ref{fig:1D}(b). 
We set the drive conditions of all atoms at P$_\mathrm{sw}$ of Fig.~\ref{fig:dse}(b).
All atomic qubits are in arbitrary states initially, 
and an arbitrary photon qubit is input into this circuit.
Then, the input photon qubit is swapped with the atomic ones successively. 
As a result, all atomic qubits are transferred to the succeeding ones after passage of the photon. 
If desired, we can skip specific atoms in this domino 
by switching off their drive fields [P$_\mathrm{id}$ of Fig.~\ref{fig:dse}(b)]. 
As another example, we present the circuit diagram of 
atom-atom $\sqrt{\rm SWAP}$ gate in Fig.~\ref{fig:1D}(c).
We set the drive conditions of atoms~1 to 4 at P$_\mathrm{sw}$, 
P$_\mathrm{id}$, P$_\mathrm{rs1/2}$, and P$_\mathrm{sw}$, respectively.
All atomic qubits are in arbitrary states initially, 
and an arbitrary photon qubit is input into this circuit.
Then, the input photon qubit is swapped with atom~1, skips atom~2, 
becomes entangled with atom~3, and is swapped again with atom~4. 
This results in the atom-atom $\sqrt{\rm SWAP}$ gate, where 
the input (output) qubits are atoms~1 and 3 (3 and 4):
$|\tone,\tone\ra_{13} \to |\tone,\tone\ra_{34}$,
$|\tone,\ttwo\ra_{13} \to 
\frac{1\mp i}{2}|\tone,\ttwo\ra_{34}+\frac{1\pm i}{2}|\ttwo,\tone\ra_{34}$,  
$|\ttwo,\tone\ra_{13} \to 
\frac{1\pm i}{2}|\tone,\ttwo\ra_{34}+\frac{1\mp i}{2}|\ttwo,\tone\ra_{34}$, and 
$|\ttwo,\ttwo\ra_{13} \to |\ttwo,\ttwo\ra_{34}$. 
The initial qubits of photon and atom~4
are transfered to atom~1 and photon, respectively, and atom~2 remains unchanged.

The proposed quantum network has the following distinct advantages.
(i)~The gate type can be controlled {\it in situ} through the atomic drive, 
without changing the circuit configuration nor 
the carrier frequencies $\om^*_l$ and $\om^*_h$ of the photonic qubits. 
(ii)~As a source of input photons, a monochromatic 
single-photon generator at $\om^*_l$ or $\om^*_h$ is sufficient, 
since this photon can be reset to a desired state 
by swapping with atom~1.
(iii)~One can skip arbitrary atoms in the circuit
by switching off their drive fields [for example, atom~2 in Fig.~\ref{fig:1D}(c)].
This implies the possibility of two-qubit gates between non-neighboring qubits,
which would substantially simplify the gate-based quantum computation.
(iv)~One-qubit gate operations to individual atoms are readily performable 
through the drive fields. % ports without perturbing other atoms. 
Therefore, combined with the $\sqrt{\rm SWAP}$ gate, 
the universal gate set is completed in the atomic network. 
We can perform universal quantum computation % in principle
by inputting microwave photons successively 
and varying the drive conditions.

In summary, we theoretically proposed a two-qubit gate 
between a superconducting atom and a propagating microwave photon. 
The gate operation completes deterministically upon reflection of the photon, 
and various two-qubit gates (including SWAP, $\sqrt{\rm SWAP}$, and Identity) 
are realizable through {\it in situ} control of the drive field. 
We can construct a quantum network of superconducting atoms
aided by microwave photons,
in which two-qubit gates are performable between non-neighboring atoms. 
This would widen the potential of superconducting quantum computing.

This work was partly supported by JSPS KAKENHI 
(Grants No. 16K05497, No. 26220601, and No. 15K17731). 

\appendix
%%%%%%%%%%%%%%%%%%%%%%%%%%%%%%%%%%%%%%%%%%%%%%%%%%%%%%%%%%%%%%%%%%%%%%%%%%%%%%%
\section{Derivation of $\xi_{ij}(\om)$}\label{sec:A}
%%%%%%%%%%%%%%%%%%%%%%%%%%%%%%%%%%%%%%%%%%%%%%%%%%%%%%%%%%%%%%%%%%%%%%%%%%%%%%%
Here, we derive the coefficients $\xi_{ij}(\om)$ which appear in Eqs.~(\ref{eq:xi11})--(\ref{eq:xi22}). 
The Hamiltonian of the overall system including waveguide~1 is given by 
\bea
\cH &=& \cH_{ar}+\cH_{rw},
\\
\cH_{ar} &=& \om_r a^{\dag}a\sig\sig^{\dag} + 
[(\om_q-\om_d)+(\om_r-2\chi)a^{\dag}a]\sig^{\dag}\sig
+\Om_d(\sig^{\dag}+\sig),
\\
\cH_{rw} &=& \int dk \left[
ka_k^{\dag}a_k + \sqrt{\kap/2\pi}(a^{\dag}a_k+a_k^{\dag}a)
\right],
\eea
where $\cH_{ar}$ describes the driven atom-resonator system [Eq.~(\ref{eq:Har})], 
$\cH_{rw}$ describes the interaction between the resonator and the propagating photon in waveguide~1,
and $a_k$ is the annihilation operator of the waveguide photon with wave number $k$. 
The superconducting atom is assumed to have an infinite lifetime here. 
Switching to the dressed-state basis [Eqs.~(\ref{eq:tone})--(\ref{eq:tfour})], $\cH$ is rewritten as
\bea
\cH &=& \sum_j \tom_j \sig_{jj}
+ \int dk \left[ ka_k^{\dag}a_k + 
\textstyle{\sum_{i,j}}
(\eta_{ji}\sig_{ji}a_k + \eta^*_{ji}a_k^{\dag}\sig_{ij})/\sqrt{2\pi}
\right],
\label{eq:Hall}
\eea
where 
the indices run over $i,j=1,\cdots,4$ % $1 \leq i,j \leq 4$ 
and $\sig_{ji}=|\widetilde{j}\ra\la \widetilde{i}|$. 
$\eta_{ji}$ is given by 
$\eta_{32}=\eta_{41}=\sqrt{\kap}\cos \theta_t$, 
$\eta_{42}=-\eta_{31}=\sqrt{\kap}\sin \theta_t$, 
and $\eta_{ji}=0$ otherwise.

We introduce the real-space representation of the field operator by
$a_r = (2\pi)^{-1/2} \int dk \ e^{ikr}a_k$. 
In this representation, the $r<0$ ($r>0$) region 
corresponds to the incoming (outgoing) field. 
From Eq.~(\ref{eq:Hall}), we can rigorously derive 
the following input-output relation, %~\cite{WM}, 
\beq
a_r(t) = a_{r-t}(0) -i\theta(r)\theta(t-r) \sum_{i,j} \eta^*_{ji}\sig_{ij}(t-r),
\label{eq:inout}
\eeq
where $\theta(r)$ is the Heaviside step function. 
We can also derive the following Heisenberg equations,
\bea
\frac{d}{dt}\sig_{13} &=& 
\left(-i\tom_{31}-\kap/2 \right)\sig_{13}
% -\frac{\eta_{31}\eta^*_{41}+\eta_{32}\eta^*_{42}}{2}\sig_{14}
+i[\eta_{31}(\sig_{33}-\sig_{11})-\eta_{32}\sig_{12}+\eta_{41}\sig_{43}]a_{-t}(0),
\\
\frac{d}{dt}\sig_{14} &=& 
\left(-i\tom_{41}-\kap/2 \right)\sig_{14}
% -\frac{\eta_{41}\eta^*_{31}+\eta_{42}\eta^*_{32}}{2}\sig_{13}
+i[\eta_{41}(\sig_{44}-\sig_{11})-\eta_{42}\sig_{12}+\eta_{31}\sig_{34}]a_{-t}(0),
\eea
where $\tom_{ij}=\tom_i-\tom_j$.

Hereafter, we consider a case in which the atom is in the state $|\tone\ra$ 
and a single photon with wavefunction $f(r)$ is input at the initial moment ($t=0$).
The initial and final state vectors are written as 
\bea
|\phi_{in}\ra &=& 
\int dr f(r) a_r^{\dag} |\tone\ra,
\\
|\phi_{out}\ra &=& 
e^{-i\tom_1 t} \int dr g_{11}(r,t) a_r^{\dag} |\tone\ra 
+ e^{-i\tom_2 t} \int dr g_{12}(r,t) a_r^{\dag} |\ttwo\ra,
\label{eq:phiout}
\eea
where $g_{11}(r,t)$ and $g_{12}(r,t)$ are the photon wavefunctions after reflection,
and the final moment $t$ is sufficiently large.  
The initial and final state vectors are connected by the unitary time evolution, 
$|\phi_{out}\ra = e^{-i\cH t}|\phi_{in}\ra$. 
Note that the natural time evolution of the dressed state ($e^{-i\tom_j t}$) is separated. 
For later convenience, we introduce
$s_{13}(t)=\la \tone|\sig_{13}(t)|\phi_{in}\ra$ and 
$s_{14}(t)=\la \tone|\sig_{14}(t)|\phi_{in}\ra$. 
Their equations of motion are given,
remembering that $a_{-t}(0)|\phi_{in}\ra = f(-t)|\tone\ra$
and that $|\tone\ra$ is an eigenstate of $\cH$, by
\bea
\frac{d}{dt}s_{13} &=& (-i\tom_{31}-\kap/2)s_{13}-i\eta_{31}f(-t),
\\
\frac{d}{dt}s_{14} &=& (-i\tom_{41}-\kap/2)s_{14}-i\eta_{41}f(-t). 
\eea
If the pulse length of the input photon is much larger than $\kap^{-1}$, 
we can adiabatically solve the above equations. 
Denoting the central frequency of the input photon by $\om$, 
the adiabatic solutions are given by
\bea
s_{13}(t) &=& \frac{-i\eta_{31}}{\kap/2-i(\om-\tom_{31})}f(-t),
\\
s_{14}(t) &=& \frac{-i\eta_{41}}{\kap/2-i(\om-\tom_{41})}f(-t).
\eea
From Eq.~(\ref{eq:phiout}), we have
$g_{11}(r,t) = e^{i\tom_1 t} \la \tone|a_r|\phi_{out}\ra = \la \tone|a_r(t)|\phi_{in}\ra$. 
Substituting Eq.~(\ref{eq:inout}) into this equation, we obtain
\beq
\xi_{11}(\om) = \frac{g_{11}(r,t)}{f(r-t)} =
1-\frac{\kap\sin^2\theta_t}{\kap/2-i(\om-\tom_{31})}
 -\frac{\kap\cos^2\theta_t}{\kap/2-i(\om-\tom_{41})}. 
\eeq
Thus, $\xi_{11}(\om)$ [Eq.~(\ref{eq:xi11})] is derived. 
$\xi_{12}$, $\xi_{21}$ and $\xi_{22}$ are derivable similarly. 

%%%%%%%%%%%%%%%%%%%%%%%%%%%%%%%%%%%%%%%%%%%%%%%%%%%%%%%%%%%%%%%%%%%%%%%%%%%%%%%
\section{averaged gate fidelity}\label{sec:B}
%%%%%%%%%%%%%%%%%%%%%%%%%%%%%%%%%%%%%%%%%%%%%%%%%%%%%%%%%%%%%%%%%%%%%%%%%%%%%%%
Here, we present the formalism for evaluation of 
the averaged gate fidelity of the atom-photon gate.
Considering the finite pulse length of the input pulse,
% and the finite lifetime of the atom. 
the input state vectors are written as
\bea
|\tone,\om_l^*\ra_{\rm in} &=& \int d\om f_{\om_l^*}(\om)|\tone,\om\ra,
\\
|\tone,\om_h^*\ra_{\rm in} &=& \int d\om f_{\om_h^*}(\om)|\tone,\om\ra,
\\
|\ttwo,\om_l^*\ra_{\rm in} &=& \int d\om f_{\om_l^*}(\om)|\ttwo,\om\ra,
\\
|\ttwo,\om_h^*\ra_{\rm in} &=& \int d\om f_{\om_h^*}(\om)|\ttwo,\om\ra,
\eea
where $f_{\om_l^*}(\om)$ is the wavefunction 
of the input photon in the frequency space. 
It is given, as the Fourier transform of Eq.~(\ref{eq:foml}) with $\om=\om_l^*$, by
\beq
f_{\om_l^*}(\om) = \sqrt{\frac{4\pi}{l^3}}
\frac{1}{(\pi/l)^2-(\om-\om_l^*)^2}\cos[(\om-\om_l^*)l/2], 
\eeq
where $l$ denotes the pulse length. 
$f_{\om_h^*}(\om)$ is defined similarly.

After reflection of the input photon, 
the state vectors evolve as Eqs.~(10)--(11). 
We also consider here the decay of the atomic excited state $|e\ra$
during the gate time $t_g$. 
Using Eqs.~(\ref{eq:tone}) and (\ref{eq:ttwo}), and denoting the atomic lifetime by $T_1$, 
the dressed states $|\tone\ra$ and $|\ttwo\ra$ evolve as
\bea
|\tone\ra & \to & |\tone'\ra = 
\cos\theta_l|g,0\ra - e^{-t_g/2T_1} \sin\theta_l|e,0\ra + \cdots,
\\
|\ttwo\ra & \to & |\ttwo'\ra = 
\sin\theta_l|g,0\ra + e^{-t_g/2T_1} \cos\theta_l|e,0\ra + \cdots,
\eea
where the dots denote the decayed states, 
which are entangled with the environment
and are out of the considered Hilbert space. 
Omitting the phase factor due to natural evolution, 
the input state vectors evolve as
\bea
|\tone,\om_l^*\ra_{\rm in} & \to & |\tone,\om_l^*\ra_{\rm out}=
\int d\om f_{\om_l^*}(\om) \xi_{11}(\om) |\tone',\om\ra
+ \cdots,
\\
|\tone,\om_h^*\ra_{\rm in} & \to & |\tone,\om_h^*\ra_{\rm out}=
\int d\om f_{\om_h^*}(\om) \xi_{11}(\om) |\tone',\om\ra + 
\int d\om f_{\om_h^*}(\om) \xi_{12}(\om) |\ttwo',\om-\Delta\om\ra,
\\
|\ttwo,\om_l^*\ra_{\rm in} & \to & |\ttwo,\om_l^*\ra_{\rm out}=
\int d\om f_{\om_l^*}(\om) \xi_{21}(\om) |\tone',\om+\Delta\om\ra + 
\int d\om f_{\om_l^*}(\om) \xi_{22}(\om) |\ttwo',\om\ra,
\\
|\ttwo,\om_h^*\ra_{\rm in} & \to & |\ttwo,\om_h^*\ra_{\rm out}=
\int d\om f_{\om_h^*}(\om) \xi_{22}(\om) |\ttwo',\om\ra
+ \cdots,
\eea
where the dots denote irrelevant terms 
that are out of the considered Hilbert space. % of the gate operation. 

On the other hand, the ideal time evolution of the SWAP gate is
\bea
|\tone,\om_l^*\ra_{\rm in} & \to & 
|\tone,\om_l^*\ra_{\rm out}^{\rm id}= |\tone,\om_l^*\ra_{\rm in}
= \int d\om f_{\om_l^*}(\om) |\tone,\om\ra, 
\label{eq:id1}
\\
|\tone,\om_h^*\ra_{\rm in} & \to & 
|\tone,\om_h^*\ra_{\rm out}= |\ttwo,\om_l^*\ra_{\rm in}
= \int d\om f_{\om_l^*}(\om)|\ttwo,\om\ra,
\\
|\ttwo,\om_l^*\ra_{\rm in} & \to & 
|\ttwo,\om_l^*\ra_{\rm out}=|\tone,\om_h^*\ra_{\rm in}
= \int d\om f_{\om_h^*}(\om)|\tone,\om\ra,
\\
|\ttwo,\om_h^*\ra_{\rm in} & \to & 
|\ttwo,\om_h^*\ra_{\rm out}=|\ttwo,\om_h^*\ra_{\rm in}
= \int d\om f_{\om_h^*}(\om)|\ttwo,\om\ra. 
\label{eq:id2}
\eea
The entanglement fidelity is given by
$f_{\rm sw}=|_{\rm out}^{\rm \ id}\la\tone,\om_l^*|\tone,\om_l^*\ra_{\rm out}
+ \cdots
+_{\rm out}^{\rm \ id}\la\ttwo,\om_h^*|\ttwo,\om_h^*\ra_{\rm out}|^2/16$,
and the averaged gate fidelity of SWAP gate is given by
$\cF_{\rm sw}=(4f_{\rm sw}+1)/5$~\cite{fid}. 
The fidelities of the other gates are obtained by 
replacing the right-hand sides of Eqs.~(\ref{eq:id1})--(\ref{eq:id2}) properly.

%%%%%%%%%%%%%%%%%%%%%%%%%%%%%%%%%%%%%%%%%%%%%%%%%%%%%%%%%%%%%%%%%%%%%%%%%%%%%

%%%%%%%%%%%%%%%%%%%%%%%%%%%%%%%%%%%%%%%%%%%%%%%%%%%%%%%%%%%%%%%%%%%%%%%%%%%%%

%%%%%%%%%%%%%%%%%%%%%%%%%%%%%%%%%%%%%%%%%%%%%%%%%%%%%%%%%%%%%%%%%%%%%%%%%%%%%%
%%%%%%%%%%%%%%%%%%%%%%%%%%%%%%%%%%%%%%%%%%%%%%%%%%%%%%%%%%%%%%%%%%%%%%%%%%%%%%
\end{document}